\documentclass[showpacs,preprintnumbers,prd,amssymb]{revtex4}
\begin{document}

\title{Lessons of spin and torsion: Reply to ``Consistent coupling to
Dirac fields in teleparallelism"}
\author{Yu.N. Obukhov}
\affiliation{Institute for Theoretical Physics, University of Cologne,
50923 K\"oln, Germany}
\altaffiliation[On leave from: ]{Department of Theoretical Physics, 
Moscow State University, 117234 Moscow, Russia}
\author{J.G. Pereira}
\affiliation{Instituto de F\'{\i}sica Te\'orica,Universidade Estadual 
Paulista\\Rua Pamplona 145,01405-900 S\~ao Paulo SP, Brazil}

\begin{abstract}
In reply to the criticism made by Mielke in the pereceding Comment 
[Phys. Rev. D69 (2004) 128501] on our recent paper, we once again explicitly 
demonstrate the inconsistency of the coupling of a Dirac field to gravitation 
in the teleparallel equivalent of general relativity. Moreover, we stress that
the mentioned inconsistency is generic for {\it all} sources with spin and 
is by no means restricted to the Dirac field. In this sense the 
$SL(4,R)$-covariant generalization of the spinor fields in the teleparallel 
gravity theory is irrelevant to the inconsistency problem.
\end{abstract}
\pacs{04.50.+h; 04.20.Jb; 03.50.Kk}
\maketitle

\section{Introduction}

The problem of coupling sources of spinning matter to the teleparallel 
gravitational field is well known, see \cite{HS,jose1,jose2,jose3}, for
example. This difficulty is naturally related to the fact that teleparallelism 
can be consistently treated as the gauge theory of the translation group of
spacetime. The corresponding dynamical current, for the generators of 
translations, is the energy-momentum. Accordingly, the teleparallelism is 
perfectly equivalent to the general relativity theory for the matter sources 
without spin. However, as it is well known, the spin current corresponds to 
the generators of the Lorentz group and in this sense it does not formally
``fit" into the gauge approach based on the group of translations. This is 
different from the more general gauge theory based on the Poincar\'e symmetry 
group (semidirect product of translations times the Lorentz group) in which 
the energy-momentum and the spin currents have equal ``rights", and in which 
they are consistently coupled to the curvature and torsion of spacetime. 

In our recent paper \cite{tele}, we have developed a metric-affine approach 
to teleparallel gravity in which the latter was treated as a particular 
case of general metric-affine gravity (MAG) specified by the geometric 
constraints of vanishing of curvature and nonmetricity. Among other results,
such an approach has provided an explicit demonstration of the inconsistency 
of the coupling of matter fields with spin. The author of the Comment 
\cite{mievas} tries to dispute this result. In our reply we will show that 
their claim is misleading. 

\section{Inconsistency of the spin coupling}\label{3fold}

Since this point seems to be a source of constant misunderstandings in the
studies of the teleparallel gravity, we will clarify the corresponding result 
by using three different techniques. 

\subsection{Tetrad approach}

At first, let us recall that one can deal with teleparallelism in the purely 
tetrad framework by taking the coframe components $h^\alpha_i$ as the basic 
field variables \cite{indices} and treating the torsion tensor $T^k{}_{ij}
= h^k_\alpha\left(\partial_ih^\alpha_j - \partial_jh^\alpha_i\right)$ as the 
translational gauge field strength \cite{jose1,jose2}. The action, with 
Yang-Mills type of Lagrangian, is 
\begin{equation}S = \int d^4x\,h\left(\frac{1}{4\kappa}\,S_k{}^{ij}
\,T^k{}_{ij} + L^{\rm mat}\right),
\end{equation}
where $h = {\rm det}\,h^\alpha_i$ and $S^{kij} = {\frac 14}\,T^{kij} 
+ {\frac 14}\,(T^{ikj} - T^{jki}) - {\frac 12}\,(g^{kj}\,T^{li}{}_l 
- g^{ki}\,T^{lj}{}_l)$. The variation of the action with respect to the 
tetrad yields the field equation
\begin{equation}\label{teleq}
{\frac 1 h}\,\partial_j(h S_k{}^{ij}) + S_l{}^{mi}\,T^l{}_{mk} - {\frac 14}
\,\delta^i_k\,S_l{}^{mn}\,T^l{}_{mn} = \kappa T_k{}^i.
\end{equation}
The right-hand side is the canonical energy-momentum of matter defined by 
the variational derivative $T_k{}^i = h_k^\alpha\,\delta (hL^{\rm mat})/(h
\delta h^\alpha{}_i)$. 

The inconsistency of the coupling of matter with spin arises as follows: The 
tetrad $h^\alpha_i$ has 16 independent components and, accordingly, the 
equation (\ref{teleq}) has also 16 components. However, there is a well-known 
geometric identity which relates the left-hand side to the Einstein tensor:
\begin{equation}
{\frac 1 h}\,\partial_j(h S_k{}^{ij}) + S_l{}^{mi}\,T^l{}_{mk} - {\frac 14}
\,\delta^i_k\,S_l{}^{mn}\,T^l{}_{mn} \equiv \widetilde{G}_k{}^i.
\end{equation}
The tilde denotes the purely Riemannian object constructed from the spacetime 
metric $g_{ij}$. Since the Einstein tensor is symmetric, $\widetilde{G}_{ij}= 
\widetilde{G}_{ji}$, we immediately discover that the field equation 
(\ref{teleq}) yields the vanishing of the antisymmetric part of the canonical 
energy-momentum tensor: $T_{[ij]} = 0$. Using the Noether conservation law of 
total angular momentum, we then find that the spin tensor $\tau^k{}_{ij} =
- \tau^k{}_{ji}$ must be conserved for itself, $\nabla_k\,\tau^k{}_{ij} = 0$. 

\subsection{MAG approach: first field equation}

The same result can be rederived within the framework of the MAG approach.
Since that was the subject of our previous paper \cite{tele}, we merely
state here that the {\it first} field equation (derived from the variation
with respect to the coframe) reads, for the teleparallel equivalent Lagrangian,
\begin{equation}
{\frac 1 2}\,\eta_{\alpha\mu\nu}\wedge\widetilde{R}^{\mu\nu} = \kappa
\Sigma_\alpha.\label{first}
\end{equation}
Here $\widetilde{R}^{\mu\nu}$ is the 2-form of the Riemannian curvature, and 
$\Sigma_\alpha$ is the 3-form of the canonical energy-momentum of matter. It 
is easy to verify that 
\begin{equation}\vartheta_{[\alpha}\wedge\eta_{\beta]\mu\nu} \equiv 
-\,\eta_{\alpha\beta[\mu}\wedge\vartheta_{\nu]}. \label{id}
\end{equation}
As a result, we straightforwardly see that the antisymmetric part of the 
left-hand side of the equation (\ref{first}) vanishes $\vartheta_{[\alpha}
\wedge\eta_{\beta]\mu\nu}\wedge\widetilde{R}^{\mu\nu} = - \eta_{\alpha\beta\mu}
\wedge\vartheta_{\nu}\wedge\widetilde{R}^{\mu\nu} \equiv 0$ in view of the 
Ricci identity $-\widetilde{R}_\mu{}^\nu\wedge\vartheta_\nu = \widetilde{D}
\widetilde{D}\vartheta_\mu = 0$. 

Consequently, we again find that the 
antisymmetric part of the energy-momentum current must vanish, $\vartheta_{
[\alpha}\wedge\Sigma_{\beta]} = 0$, and hence the spin two-form $\tau_{
\alpha\beta} = \tau^k{}_{\alpha\beta}\,\eta_k$ should be conserved: 
$D\,\tau_{\alpha\beta} = 0$.

\subsection{MAG approach: second field equation}

Finally, let us prove the above result by following the same reasoning of 
the author of the Comment \cite{mievas} who analyzed the {\it second} 
field equation. It reads (see eq. (4.3) of \cite{mievas}):
\begin{equation}
D\,\lambda_{\alpha\beta} + \vartheta_{[\alpha}\wedge H^{||}_{\beta]} 
= \tau_{\alpha\beta}.\label{second}
\end{equation}
Since the teleparallelism equivalent translational momentum is given by
$H^{||}_\alpha = (1/\ell^2)\,\eta_{\alpha\mu\nu}\wedge K^{\mu\nu}$ in terms 
of the contortion 1-form $K^{\mu\nu}$, we have
\begin{equation}
\vartheta_{[\alpha}\wedge H^{||}_{\beta]} \equiv -\,{\frac 1 {\ell^2}}
\,D\,\eta_{\alpha\beta}.\label{id2}
\end{equation}
Indeed, using the identity (\ref{id}), we find $\vartheta_{[\alpha}\wedge
\eta_{\beta]\mu\nu}\wedge K^{\mu\nu} = - \eta_{\alpha\beta\mu}\wedge 
\vartheta_\nu\wedge K^{\mu\nu} = - \eta_{\alpha\beta\mu}\wedge T^\mu$. Then
the above result is easily found with the help of the general formulas 
(3.8.5) of \cite{PR} which give the covariant derivatives of the $\eta$-forms.
Substituting (\ref{id2}) into (\ref{second}) and by subsequently taking the
covariant exterior derivative, we obtain $D \,\tau_{\alpha\beta}=0$, since
$DD(\lambda_{\alpha\beta}  - \eta_{\alpha\beta}/\ell^2) = 0$ in view of the
teleparallel constraint requiring the vanishing of the total curvature. 

Thus, Eq.~(4.6) of the Comment \cite{mievas} is totally misleading in the 
sense that a clear zero is ``hidden" in the second term on the left-hand side. 

\section{Making spin coupling consistent}

We have demonstrated above that the gravitational coupling of spin is 
generically inconsistent in the teleparallel equivalent gravity. How can one
cure this situation? The source of the difficulty is clear: the left-hand 
side (geometric one) of the gravitational field equation is symmetric, whereas 
the right-hand side (source) is asymmetric for matter with spin. 
Correspondingly, one can proceed in one of the two ways: (i) introduce a 
different coupling rule so that the energy-momentum becomes symmetric, or 
(ii) change the dynamical scheme so that the geometric left-hand side 
also becomes asymmetric. 

\subsection{Alternative coupling prescription: Einstein's theory}

In \cite{jose3}, and more recently in \cite{maluf} (see also the earlier
discussion in \cite{HS}), it was noticed that if the coupling Lagrangian of a 
spinor field contains not the Weitzenb\"ock connection of the teleparallelism,
but the usual Riemannian connection, then the coupling inconsistency 
disappears. In this case, the equation (\ref{first}) is replaced by
\begin{equation}
{\frac 1 2}\,\eta_{\alpha\mu\nu}\wedge\widetilde{R}^{\mu\nu} = \kappa
\,\sigma_\alpha.\label{firstBR}
\end{equation}
The 3-form $\sigma_\alpha$ on the right-hand side is the so-called
Belinfante-Rosenfeld energy-momentum. It is symmetric, $\vartheta_{[\alpha}
\wedge\sigma_{\beta]} = 0$. Consequently, there is not any coupling
inconsistency. The teleparallel gravity with such a coupling prescription
becomes  indistinguishible from Einstein's general relativity theory.

\subsection{From translations to Poincar\'e group: Einstein-Cartan theory}

An alternative procedure is to include the spin, together with the 
energy-momentum current, as a dynamical source of equal right for the 
gravitational field. This naturally leads to the gauge theory based on the 
Poincar\'e symmetry group with the generators of translations related to the 
canonical energy-momentum $\Sigma_\alpha$ and the generators of the Lorentz 
group related to spin $\tau_{\alpha\beta}$. Such an extension of the dynamical 
contents yields an extension of the spacetime geometry to the Riemann-Cartan 
case with nontrivial curvature $R_\alpha{}^\beta$ and torsion $T^\alpha$ 
2-forms. The extended Einstein-Hilbert Lagrangian then yields the 
Einstein-Cartan field equations
\begin{eqnarray}
{\frac 1 2}\,\eta_{\alpha\mu\nu}\wedge R^{\mu\nu} &=& \kappa\,\Sigma_\alpha,
\label{firstEC}\\ 
\label{secondEC}{\frac 1 2}\,\eta_{\alpha\beta\mu}\wedge T^\mu &=& \kappa
\,\tau_{\alpha\beta}.
\end{eqnarray}
This system is completely consistent in the sense that the antisymmetric part 
of the Eq.~(\ref{firstEC}), combined with the second equation (\ref{secondEC}),
yields the Noether identity $D\tau_{\alpha\beta} + \vartheta_{[\alpha}\wedge
\Sigma_{\beta]} = 0$. 

\section{Discussion}

In the recent Comment to our paper \cite{mievas} it has been claimed that 
the spinor field couples consistently to the teleparallel gravitational field. 
However, this is incorrect and we have presented at least three direct
demonstrations of the inconsistency of the spin coupling in the teleparallel
equivalent gravity model. 

1) In simple terms, the mentioned inconsistency 
arises from the fact that the left-hand side of the teleparallel gravitational 
field equation is symmetric,whereas the right-hand side is represented by the 
canonical energy-momentum which is nonsymmetric for the matter with spin. 

2) The inconsistency is {\it generic}, i.e., it is not specific for the Dirac 
spinor field, but rather concerns all sources with spin. In this sense, the 
remark of \cite{mievas} about the necessity of considering more general 
$SL(4,R)$-covariant multispinors is irrelevant. The spin coupling 
inconsistency will be present for such matter as well.

3) The well-known fact that the teleparallel equivalent Lagrangian differs 
from the Einstein-Hilbert Lagrangian by a total derivative \cite{PR,eff}
is also irrelevant for the demonstration of the coupling inconsistency. 
Neither in our original paper \cite{tele}, nor in the derivations above 
did we ever need or use that fact. 

We have shown that, contrary to the erroneous statement of \cite{mievas}, 
the coupling of spin can only be made consistent either by the change of the
coupling prescription (thereby formally obtaining a description equivalent 
to the general relativity) or by the change of the dynamical scheme (thus 
arriving at the Einstein-Cartan gravity theory). 

\begin{acknowledgments}
The work of YNO was supported by the Deutsche Forschungsgemeinschaft (Bonn) 
with the grant HE~528/20-1. JGP thanks FAPESP and CNPq for partial financial 
support. We are grateful to Friedrich Hehl for helpful comments.
\end{acknowledgments}


\begin{thebibliography}{30}
\bibitem{indices}
Our notations and conventions are as in \cite{tele,PR}. In particular,
the Greek indices $\alpha,\beta,\dots$ label the anholonomic frame
components ($\vartheta^\alpha$, e.g.), and the Latin indices $i,j,\dots$ 
label the local spacetime coordinates ($x^i$ and $dx^i$, for example).
\end{thebibliography}
\end{document}